\documentclass[draft, preprintnumbers, nofootinbib, preprint, 12pt]{revtex4}
\usepackage{amsmath,amssymb,bm}
\begin{document}
{~}
\vspace{3cm}

\title{
Kaluza-Klein Multi-Black Holes in Five-Dimensional \\
Einstein-Maxwell Theory
\vspace{1cm}
}
\author{
	Hideki Ishihara\footnote{E-mail:ishihara@sci.osaka-cu.ac.jp}, 
	Masashi Kimura\footnote{E-mail:mkimura@sci.osaka-cu.ac.jp}, 
  	Ken Matsuno\footnote{E-mail:matsuno@sci.osaka-cu.ac.jp},
and
	Shinya Tomizawa\footnote{E-mail:tomizawa@sci.osaka-cu.ac.jp}	
}
\affiliation{ 
Department of Mathematics and Physics,
Graduate School of Science, Osaka City University,
3-3-138 Sugimoto, Sumiyoshi, Osaka 558-8585, Japan
\vspace{3cm}
}

\begin{abstract}
We construct Kaluza-Klein multi-black hole solutions on the Gibbons-Hawking
multi-instanton space in the five-dimensional Einstein-Maxwell theory.
We study geometric properties of the multi-black hole solutions. In particular, unlike
the Gibbons-Hawking multi-instanton solutions, each nut-charge is able to
take a different value due to the existence of black hole on it.
The spatial cross section of each horizon is admitted to have the topology of a different lens space  $L(n;1)=$S$^3/{\mathbb Z}_{n}$ in addition to S$^3$.
\end{abstract}

\preprint{OCU-PHYS 246}
\preprint{AP-GR 33}

\date{\today}
\maketitle

\section{Introduction}

It is well known that the Majumdar-Papapetrou solution~\cite{M-P}
describe an arbitrary number of extremely charged static black holes
in four-dimensions. The construction of the solution is possible 
because of a force balance between the gravitational
and Coulomb forces.
These solutions have been extended to multi-black holes in arbitrary
higher dimensions~\cite{Myers}, and to multi-black $p$-branes~\cite{GHT}.
All multi-black objects are constructed on the Euclid space which is transverse to the objects. The metric and the gauge potential
one-form are given by a solution 
of the Laplace equation in the Euclid space with point sources. 

In the context of supersymmetry, five-dimensional Einstein-Maxwell theory
gathers much attention. 
It is possible to construct black hole solutions 
by using the four-dimensional Euclidean self-dual 
Taub-NUT space~\cite{GGHPR, EEMRIM, GSY, IM} 
instead of the Euclid space.
The black hole solutions on the Taub-NUT space have 
S$^3$ horizons, 
and have an asymptotic structure of four-dimensional 
locally flat spacetime with 
a twisted S$^1$. 
Such black hole solutions associated with a compact dimension,
so-called Kaluza-Klein black holes, are interesting because
the solutions would connect the higher-dimensional spacetime
with the usual four-dimensional world.
 
Actually, 
if the Euclid space is replaced with any Ricci flat space,
we can obtain solutions to the Einstein-Maxwell equations 
by use of a harmonic function on the space. 
Indeed, since the Laplace equation is linear, we can
construct solutions describing multi-object by superposition of
harmonic functions with a point source. 
However, it should be clarified whether the objects
are really black holes, 
i.e., all singularities are hidden behind horizons.

In this article, as the generalization of the Kalzua-Klein black hole solutions~~\cite{GGHPR, EEMRIM, GSY, IM} to multi-black hole solutions,
we construct Kaluza-Klein multi-black holes on the Gibbons-Hawking
multi-instanton space explicitly which are solutions to the five-dimensional Einstein-Maxwell system. 
Then, we study geometric properties of the multi-black hole solutions.

\section{Construction of solutions}
\subsection{A single extreme black hole}

We start with the five-dimensional Einstein-Maxwell system 
with the action :
\begin{equation}
 S=\frac{1}{16\pi G}\int d^5x \sqrt{-g}\left( R - F_{\mu\nu}F^{\mu\nu} \right),
\end{equation}
where $G,~R$ and $\bm{F} = d \bm{A} $ 
are the five-dimensional gravitational constant, 
the Ricci scalar curvature 
and the Maxwell field with the gauge potential $\bm A$. 
{}From this action, we write down the Einstein equation 
\begin{equation}
 R_{\mu\nu}-\frac{1}{2} R g _{\mu\nu} 
		= 2 \left( F_{\mu\lambda} F^{~\lambda}_{\nu} 
 - \frac{1}{4} g _{\mu\nu} F_{\alpha \beta} F^{\alpha \beta} \right),           \label{Ein}
\end{equation}
and the Maxwell equation
\begin{equation} 
 F^{\mu\nu}_{~~~;\nu} =0. \label{Max}
\end{equation}

The extreme Kaluza-Klein black hole~\cite{GGHPR, GSY, EEMRIM, IM} 
is an exact solution of Eqs.\eqref{Ein} and \eqref{Max} constructed 
on the Taub-NUT space. 
The metric and the gauge potential one-form are written as 
\begin{eqnarray}
 ds^2 &=& - H^{-2} dT^2 + H ds^2_{\text{TN}},
        \label{SUSYBH} \\
    A &=& \pm \frac{\sqrt{3}}{2} H^{-1} dT.
\label{gaugepotintial}
\end{eqnarray}
When the Taub-NUT space is described in the form 
\begin{eqnarray}
 &&ds^2_{\text{TN}} 
		= V^{-1} \left( dR^2 + R^2d\Omega_{S^2}^2 \right)  
		+V \left( d\zeta+ N\cos\theta d\phi\right) ^2, \label{TN} \\
 &&d\Omega_{S^2}^2 = d\theta ^2 + \sin ^2 \theta d\phi^2, \\
 &&V^{-1} ( R ) = 1 + \frac{ N }{ R },                                          \label{HarmonicW}
\end{eqnarray}
with 
\begin{eqnarray}
 0\leq \theta \leq \pi,\quad 0\leq \phi \leq 2\pi,\quad 
		0\leq \zeta \leq 2\pi L ,
\label{period}
\end{eqnarray}
the function $H$ is given by
\begin{eqnarray}
 H(R) &=&  1 + \frac{ M }{ R },                                                      \label{HarmonicH}
\end{eqnarray} 
where $L, M$ and $N$ are positive constants. 
Regularity of the spacetime requires that 
the nut charge, $N$, and the asymptotic radius of S$^1$ along 
$\zeta$, $L$, are related by 
\begin{equation}
 N = \frac{ L }{ 2 } n,
\end{equation} 
where $n$ is an natural number. 

We can see that a degenerated horizon exists at $R=0$, 
where the nut singularity of the Taub-NUT space were located. 
Although the Taub-NUT space is regular only when $n=1$ 
on the conditions \eqref{period}, 
the nut singularity with $n\geq 2$ is resolved by the 
event horizon in the black hole solution~\cite{EEMRIM}.

In the case of $n=1$, 
the horizon has the shape of round S$^3$ 
in a static time-slice in contrast to the non-degenerated case, where
the horizon is squashed~\cite{IM}. 
The spacetime is asymptotically locally flat, i.e., 
a constant S$^1$ fiber bundle over the four-dimensional 
Minkowski spacetime at $R\rightarrow \infty$. 
Therefore, the spacetime behaves as a five-dimensional 
black hole near horizon, while the dimensional reduction to four-dimension 
is realized in a far region. 
In the case of $n\geq 2$, the horizon is in shape of 
lens space $L(n;1) =S^3/{\mathbb Z}_n$~\cite{EEMRIM}. 
The mass which is defined by the Komar integral at spatial infinity 
and the electric charge 
satisfy the extremality condition
\begin{eqnarray}
 M_{\rm Komar}
	= \frac{3\pi}{2G}LM 
  = \frac{\sqrt{3}}{2}|Q|. 
\label{mass}
\end{eqnarray}
The black hole solution \eqref{SUSYBH} with $n=1$ contains 
two popular spacetimes as limits: 
the five-dimensional extreme Reissner-Nordstr\"om black hole  
as $L \to \infty$ and $M\rightarrow 0$ with $LM$ finite,  
and the Gross-Perry-Sorkin monopole~\cite{GPS} as $ M \to 0 $.

\subsection{Multi-black holes}
When we generalize the single black hole solution (\ref{SUSYBH}) to 
the multi-black holes, 
it is natural to generalize the Taub-NUT space 
to the Gibbons-Hawking space~\cite{GH} 
which has multi-nut singularities. The metric form of the Gibbons-Hawking space is 
\begin{eqnarray}
& & d{s}^2_{\rm GH}
	=V^{-1} \left( d{\bm x}\cdot d{\bm x}\right)
		+V\left(d\zeta + {\bm\omega} \right)^2, 
\\
& &V^{-1} =1+\sum_i\frac{N_i}{|\bm{x}-\bm{x}_i|},
\label{042802}
\end{eqnarray}
where $\bm{x}_i=(x_i,y_i,z_i)$ denotes position of the $i$-th 
nut singularity with nut charge $N_i$ 
in the three-dimensional Euclid space, and $\bm\omega$ satisfies 
\begin{eqnarray}
{\bm \nabla}\times {\bm \omega} = {\bm\nabla} \frac{1}{V}.
\label{0428025}
\end{eqnarray}
We can write down a solution ${\bm \omega}$ explicitly as
\begin{eqnarray}
{\bm \omega}
	=\sum_i N_i 
			\frac{(z-z_i)}{|\bm{x}-\bm{x}_i|}~
			\frac{(x-x_i)dy -(y-y_i)dx}{(x-x_i)^2+(y-y_i)^2}.
\end{eqnarray}

If we assume the metric form with the Gibbons-Hawking space instead 
of the Taub-NUT space in \eqref{SUSYBH}, 
the Einstein equation (\ref{Ein}) and 
the Maxwell equation (\ref{Max}) reduce to 
\begin{equation}
	\triangle_{\rm GH} H=0, 
\label{042801}
\end{equation}
where $\triangle_{\rm GH}$ is the Laplacian of the Gibbons-Hawking space.
In general, it is difficult to solve this equation, 
but if one assume $\partial_{\zeta}$ to be a Killing vector, 
as it is for the Gibbons-Hawking space,  
then Eq.(\ref{042801}) reduces to the Laplace equation in the 
three-dimensional Euclid space, 
\begin{equation}
	\triangle_{\text E} H=0. \label{042804}
\end{equation}

We take a solution with point sources to Eq. \eqref{042804} 
as a generalization of Eq. \eqref{HarmonicH}, 
and we have the final form of the metric 
\begin{eqnarray}
&&
	ds^2=-H^{-2}dT^2 + H ds_{\rm GH}^2, 
\label{eq:ekk}
\\ 
&&H=1+\sum_i\frac{M_i}{|\bm{x}-\bm{x}_i|},
\label{eq:harmonic}
\end{eqnarray}
where  $M_i$  are constants.\footnote{
For the special case $H=1/V$ , the metric (\ref{eq:ekk}) reduces to
the four dimensional Majumdar-Papapetrou multi-black holes 
with twisted constant $S^1$, which is seen in ref.\cite{GGHPR}. 
In this special case, $M_i$ are also quantized as well 
as nut charges.
}

\section{Properties}
\subsection{Regularity}
In Eqs.\eqref{eq:ekk} with \eqref{eq:harmonic}, 
a point source labeled by ${\bm x}_i$ with $M_i >0$ and  $N_i>0$ 
is a black hole 
since the point becomes the Killing horizon and 
the cross section of it with a static time-slice has a finite area, 
as is seen below. 
A point source of $M_i >0$ and $N_i=0$, however, becomes 
a naked singularity. 
This is the reason why we need to generalize the Taub-NUT 
space to the Gibbons-Hawking space.\footnote{
Multi-black hole solutions in Taub-NUT space are constructed~\cite{EEMRIM} 
by using black ring in the space.}
For a point with $M_i < 0$ or $N_i < 0$, 
a naked singularity also appears. 
Therefore, in the case of $M_i>0$ and $N_i > 0$ for all $i$, 
the metric describes multi-black holes, 
on which we focus our attention. 

On the other hand, a point source with $M_i=0$ and $N_i>0$ 
corresponds to a Gross-Perry-Sorkin-type monopole with a 
nut charge $N_i$. 
In the case of $M_i=0$ and $N_i > 0$ for all $i$, 
the metric becomes the Gross-Perry-Sorkin multi-monopole solution~\cite{GPS}.
This multi-monopole solution is regular when 
all of the nut charges $N_i$ have to take the same value $L/2$~\cite{GH},
since the nut charges with different values from it yield the cone-singularities. 
However, the existence of black holes drastically changes this situation 
because the nut singularity with 
$N_i=L/2\times n_i (n_i :\mbox{natural number})$ converts 
into the black hole whose topology is a lens space 
$L(n_i;1)={\text S}^3/{\mathbb Z}_{n_i}$, as mentioned below. 
Therefore, all $N_i$ can take different values associated with 
the different $n_i$.

Now, we investigate the regularity on the black hole horizon. 
For simplicity, we restrict ourselves to the solution with two black holes. 
In order to examine that the geometry near the horizon ${\bm x}={\bm x}_1$, we make the coordinate transformation such that the point source ${\bm x}_1$ is 
the origin of the three-dimensional Euclid space and ${\bm x_2}=(0,0,-a)$. 
In this case, from Eqs.(\ref{eq:ekk}) and (\ref{eq:harmonic}) 
the metric can take the following simple form,
\begin{eqnarray}
	ds^2&=&-H(R,\theta)^{-2}dT^2\cr
		& &\hspace{1cm} +H(R,\theta)\biggl[
		V(R,\theta)^{-1}(dR^2+R^2d\Omega_{S^2}^2)
		+V(R,\theta) \biggl(d\zeta+\omega_\phi(R,\theta) d\phi\biggr)^2
\biggr], 
\end{eqnarray}
with
\begin{eqnarray}
&&H(R,\theta)
		=1+\frac{M_1}{R}+\frac{M_2}{\sqrt{R^2+a^2+2aR\cos\theta}},\\
&&V(R,\theta)^{-1}
		=1+\frac{N_1}{R}+\frac{N_2}{\sqrt{R^2+a^2+2aR\cos\theta}},\\
&&\omega_\phi(R,\theta)
		=N_1\cos\theta+\frac{N_2(a+R\cos\theta)}{\sqrt{R^2+a^2+2aR\cos\theta}}, 
\end{eqnarray}
where the parameter $a$ denotes the separation between 
two point sources ${\bm x}_1$ and ${\bm x}_2$ 
in the three-dimensional Euclid space. 

If we take the limit $R=|{\bm x}-{\bm x}_1|\rightarrow 0$, 
we can see the leading behavior of the metric as follows,
\begin{eqnarray}
	ds^2\simeq -\biggl(\frac{R}{M_1}\biggr)^2dT^2
		+\frac{M_1}{R}\bigg[\frac{R}{N_1}
			\biggl(d\zeta + N_1\cos\theta d\phi\biggr)^2
		+\frac{N_1}{R}\left(dR^2+R^2d\Omega_{S^2}^2\right)\biggr].
\end{eqnarray}
We should note that the other black holes does not contribute 
to this behavior of the metric in the leading order. 
Therefore, the form of each black hole is equivalent with 
the single extremal black hole \eqref{SUSYBH} 
in the vicinity of the horizon.\footnote{
Smoothness of horizon in the higher-dimensional multi-black holes 
would be lost~\cite{GHT, smooth}.}
The Kretschmann invariant near the horizon 
$R=0$ can be computed as,
\begin{eqnarray}
R_{\mu\nu\lambda\rho}R^{\mu\nu\lambda\rho}
	=\frac{19}{4M_1^2 N_1^2}+O\biggl(\frac{R}{a}\biggr),
\end{eqnarray}
which suggests the horizon $R=0$ is regular.
In fact, under the coordinate transformation,
\begin{eqnarray}
&& u = T-F(R),
\\
&&\frac{dF(R)}{dR}=\bigg(1+\frac{M_1}{R}+\frac{M_2}{a} \bigg)^{3/2}
\bigg(1+\frac{N_1}{R}+\frac{N_2}{a} \bigg)^{1/2},
\end{eqnarray}
the metric near the horizon $R=0$ has the following regular form
\begin{eqnarray}
ds^2 \simeq -2\sqrt{\frac{N_1}{M_1}}dudR 
+M_1N_1\bigg[
\biggl(\frac{d\zeta}{N_1} + \cos\theta d\phi\biggr)^2
+d\Omega_{S^2}^2\biggr].
\end{eqnarray} 
Of course, from the similar discussion, the regularity of 
the other black hole horizon ${\bm x}={\bm x}_2$ is also assured. 
Outside the black holes, there is evidently no place with a 
singular point from the explicit form of the metric. 
Even if we consider the situations with more than two black holes, 
these properties do not change in such spacetimes. 

\subsection{Geometry near horizons}

The induced metric on an intersection of the $i$-th black hole horizon 
with a static time-slice is
\begin{eqnarray}
ds^2_{{\rm Horizon}}
	=\frac{L M_i n_i}{2}\biggl[
	\biggl(\frac{d\psi}{n_i}+\cos\theta d\phi \biggr)^2
		+d\Omega_{S^2}^2\biggr],
\end{eqnarray}
where 
$0\leq \psi=2\zeta/L \leq 4\pi$.   
In the case of $n_i=1$, it is apparent that 
the $i$-th black hole is a round $S^3$, 
however, in the case of $n_i\ge 2$, the topological structure 
becomes a lens space $L(n_i;1)= {\text S}^3/{\mathbb Z}_{n_i}$.

In an asymptotically flat stationary five-dimensional black hole 
spacetime, the  only possible geometric type of 
spatial cross section of horizon is restricted to 
$S^3$ or $S^1\times S^2$~\cite{Cai,Helfgott}, 
which is the extension of Hawking's theorem on event horizon 
topology~\cite{Hawking} to five dimensions. 
However, each Kaluza-Klein black hole horizon 
can have a topological 
structure of lens spaces S$^3/{\mathbb Z}_{n_i}$ besides S$^3$. 
This fact does not contradict with the theorem in Ref.\cite{Cai,Helfgott} 
where the boundary is assumed to be asymptotically flat since the Kaluza-Klein 
black holes are not asymptotically flat.

\subsection{Asymptotic structure}
Finally, we study the asymptotic behavior of the Kaluza-Klein 
multi-black hole in the neighborhood of the spatial infinity 
$R\rightarrow\infty$. 
The functions $H$, $V^{-1}$ and 
${\bm \omega}$ behave as
\begin{eqnarray}
&&H(R,\theta)
	\simeq 1+\frac{\sum_i M_i}{R}+O\biggl(\frac{1}{R^2}\biggr),
\\
&&V(R,\theta)^{-1}
	\simeq 1+\frac{\sum_i N_i}{R}+O\biggl(\frac{1}{R^2}\biggr),
\\
&&{\bm\omega}(R,\theta)
	\simeq \biggl(\sum_i N_i\biggr)\cos\theta d\phi+O\biggl(\frac{1}{R}\biggr).
\end{eqnarray}
We can see that the spatial infinity possesses the structure of 
S$^1$ bundle over S$^2$ such that it is a 
lens space $L(\sum_i n_i;1)$. 
For an example, in the case of two Kaluza-Klein black holes 
which have the same topological structure of S$^3$, 
the asymptotic structure is  topologically  homeomorphic 
to the lens space $L(2;1)={\text S}^3/{\mathbb Z}_2$.  
{}From this behavior of the metric near the spatial infinity, 
we can compute the Komar mass at spatial infinity of this multi-black hole 
system as 
\begin{eqnarray}
	M_{\text{Komar}}=\frac{3\pi}{2G}L\sum_i M_i.
\end{eqnarray}
Since the total electric charge is given by
\begin{eqnarray}
	Q_{\text{total}}=\sum_i Q_i = \frac{\sqrt{3}\pi}{G}L\sum_i M_i, 
\end{eqnarray}
then the total Komar mass and the total electric charge satisfy
\begin{eqnarray}
	M_{\text{Komar}}= \frac{\sqrt{3}}{2} |Q_{\text{total}}|.
\end{eqnarray}
Therefore, from Eq.(\ref{mass}) we find that an observer located in the neighborhood of 
the spatial infinity feels as if there were a single 
Kaluza-Klein black hole with the point source with  
the parameter $M=\sum_i M_i$ and the nut charge 
$N={\sum_i N_i}$. 

\section{Summary and Discussion}

In conclusion, 
we constructed the Kaluza-Klein multi-black hole solutions on the nuts of the 
Gibbons-Hawking space as solutions in the five-dimensional Einstein-Maxwell theory. We also investigated the properties of these solutions, in particular, the regularity, the geometry near horizon and the asymptotic structure. In the solution Eqs.(\ref{eq:ekk}) and (\ref{eq:harmonic}) a point source labeled by ${\bm x}_i$ with $M_i>0$ and $N_i>0$ is a black hole. One of the most interesting properties is that the possible spatial topology of 
the horizon of each black hole is the lens space S$^3/{\mathbb Z_{n_i}}$, 
where the natural number $n_i$ is related with the value $N_i$ of each nut charge by $N_i=L/2\times n_i$. The spatial infinity has the structure of $S^1$ bundle over $S^2$ such that it is a lens space $L(\sum_in_i;1)$. 

This fact suggests that two black holes with $S^3$ horizons constructed in this article coalesce into a black  hole with a $L(2;1)=S^3/{\mathbb Z}_2$ horizon.
By the coalescences, black holes may change into a black hole with a different lens space $L(n;1)$. It is also expected that the area of a single black hole formed by such process would be different from that in a topology-preserving process. Observing such difference might lead to the verification of the existence of extra dimensions. Therefore, from this view point, the coalescence of black holes with the topologies of various lens spaces may be interesting physical phenomenon.

\section*{Acknowledgements}
We thank K. Nakao and Y. Yasui for useful discussions. 
This work is supported by the Grant-in-Aid
for Scientific Research No.14540275 and No.13135208.


\begin{thebibliography}{99}

\bibitem{M-P}
S.D.Majumdar, Phys. Rev.  {\bf 72}, 390 (1947); \\
A. Papapetrou, Proc. R. Ir. Acad. Sect. A 51, 191 (1947).


\bibitem{Myers}
R.~C.~Myers, Phys. Rev.  {\bf D 35}, 455 (1987).

\bibitem{GHT}
G. W. Gibbons, G. T. Horowitz, and P. K. Townsend,
Class. Quantum Grav. {\bf 12}, 297 (1995).


\bibitem{GGHPR}
J.P. Gauntlett, J.B. Gutowski, C.M. Hull, S. Pakis, and H.S. Reall, 
Class. Quantum Grav. {\bf 20} 4587, (2003).

\bibitem{EEMRIM}
H.Elvang, R. Emparan, D.Mateos, and H.S.Reall, JHEP {\bf 08}, 042 (2005).

\bibitem{GSY} 
D. Gaiotto, A. Strominger, and X. Yin, 
 JHEP {\bf 02}, 024 (2006).  

\bibitem{IM}
H.Ishihara and K.Matsuno, Prog. Theor. Phys. {\bf 116}, 417 (2006).

\bibitem{GPS}
D.J. Gross and M.J. Perry, Nucl. Phys. B {\bf 226}, 29 (1983); \\
R. Sorkin, Phys. Rev. Lett. {\bf 51}, 87 (1983).

\bibitem{GH}
G.W. Gibbons , S.W. Hawking , Phys. Lett. {\bf 78B} 430, (1978).


\bibitem{smooth}
D.L. Welch, Phys. Rev. D {\bf 52}, 985 (1995).

\bibitem{Cai}
M. I. Cai and G. J. Galloway, Class Quant. Grav. {\bf 18}, 2707 (2001).


\bibitem{Helfgott}
C. Helfgott, Y. Oz, and  Y. Yanay, JHEP 0602, 025 (2006).

\bibitem{Hawking}
S. W. Hawking, Commun. Math. Phys. {\bf 25}, 152 (1972).

\end{thebibliography}
\end{document}